\documentstyle[12pt]{article}
\begin{document}
\title{On The Quantization Of Constraint Systems:\\ A Lagrangian Approach}
\author{F. Loran\thanks{e-mail:
loran@cc.iut.ac.ir}\\ \\
  {\it Department of  Physics, Isfahan University of Technology (IUT)}\\
{\it Isfahan,  Iran,} \\
  {\it Institute for Studies in Theoretical Physics and Mathematics (IPM)}\\
{\it P. O. Box: 19395-5531, Tehran, Iran.}}
\date{}
\maketitle
\begin{abstract}
It is possible
to introduce external time dependent back ground fields in the formulation of a system
as fields whose dynamics can not be deduced from Euler Lagrange equations of motion.
This method leads to singular Lagrangians for real systems. We discuss quantization
of constraint systems in these cases and introduce generalized Gupta-Bleuler
quantization. In two examples we show explicitly that this method of quantization
leads to true Schr\"{o}dinger equations.
\end{abstract}
\section{Introduction}
Obtaining Euler Lagrange equations of motion for a system given
by a Lagrangian, $L=L(x_i,\dot{x}_i;t)$, it may happen that some
of the accelerations $\ddot{x}_i$s remain unsolved. In these
cases the corresponding coordinates become arbitrary functions of
time. These systems are called constraint systems
\cite{Dirac,Berg}. In general, constraint systems possess gauge
degrees of freedom. It is believed that gauge degrees of freedom
are not physical. They should be eliminated for example by
imposing gauge fixing conditions and defining reduced phase
space. Gauge invariance should be completely eliminated because
one deals only with gauge invariant objects \cite{Henbook}. In
other words, observables are independent of gauges. Although
eliminating gauge degrees of freedom is sufficient to fulfill the
above considerations but it is not a necessary condition and
causes some ambiguities in the general formulation of constraint
systems. Part of these ambiguities are due to the fact that in
principle it is possible to introduce external time dependent
back ground fields in the formulation of a system as fields
(coordinates) whose dynamics can not be determined by Euler
Lagrange  equations of motion (EL). Obviously such formulation
leads to singular Lagrangians and physical degrees of freedom
(may) become gauged. Consequently it is not reasonable to
eliminate them. In this article we introduce a consistent
approach for quantization by generalizing Gupta-Bleuler
quantization \cite{Itz}. The paper is organized as follows.
Section 2 is a brief review of the reduced phase space
quantization. In Section 3 we introduce generalized Gupta-Bleuler
quantization and show that it is equivalent to Schr\"{o}dinger
approach when applied to unconstrained systems given by regular
Lagrangians. In Section 4 we define reduced Hilbert space and
examine the generalized Gupta-Bleuler quantization in two
examples. Finally in section 5 we summarize our results.
\section{Reduced Phase Space Quantization}
Euler Lagrange equations of motion are not sufficient to
determine dynamic of a constraint system. In other words, given a
Lagrangian $L(x_i,\dot{x}_i)$ of a constrained system, the
Hessian matrix
$W_{ij}=\frac{\partial^2L}{\partial\dot{x}_i\partial\dot{x}_j}$ is
singular, $\det W=0$. The direct consequence of this singularity
is that, momenta conjugate to coordinates are not independent and
satisfy some relations called primary constraints (PC). The
dynamic of the system should be consistent with PCs. Here,
consistency means that these constraints should be preserved in
time. This may generate some new constraints called secondary
constraints (SC). If the system possesses gauge degrees of
freedom, one fixes the gauges by imposing new constraints known
as gauge fixing conditions. The classical trajectory of the
system lies on a subspace of the phase space, the reduced phase
space (RPS), which is the intersection of the subspaces defined
by PCs, SCs and gauge fixing conditions. In principle one can
reparameterize the phase space by new coordinates
$(X_i,X_\alpha)$, where $X_i$'s are coordinates of the RPS and
$X_\alpha$'s are the remaining. These new coordinates satisfy the
following algebra,
\begin{eqnarray}\begin{array}{ccc}
\{X_i,X_j\}^*=\sigma_{ij},&\{X_i,X_\alpha\}^*=0,&\{X_\alpha,X_\beta\}^*=0,
\end{array}\end{eqnarray}
where $\{\ ,\ \}^*$ is the Dirac bracket and $\sigma_{ij}$ is a
nontrivial two form. The classical equations of motion are given
by the Hamilton equation, $\dot{F}(X_i)=\{F(X_i),H\}^*$, in which
$H=H(X_i)$ is the well defined Hamiltonian on the RPS. For
quantization one replaces Dirac brackets with commutators,
\begin{equation}
\{X_i,X_j\}^*=\sigma_{ij}\to[\hat{X}_i,\hat{X_j}]=i\hbar\sigma_{i,j},
\end{equation}
where $\hat{X}_i$ are operators corresponding to coordinates
$X_i$ and dynamic is given by the Hamiltonian
$\hat{H}=H(\hat{X}_i)$. This is the well known Dirac quantiztion.
In this article we call it RPS quantization to remind its
structure. Most of the other methods for quantization of
constrained systems follow the same ideas as those in the RPS
method \cite{Henbook, Gitman, Gov, Klauder}. For example in the
Faddeev-Popov method the measure of path integral is written such
that the path integral formulation of constrained systems turns
out to be similar to the formulation of unconstrained systems on
the RPS.
\section{Generalized Gupta-Bleuler Quantization}
Although PCs are generated by definition of momenta but SCs have
their roots in dynamic. They can be (partly) obtained from Eular
Lagrange equations of motion \cite{Pons}. It is natural to ask
why these classical equations of motion should be satisfied in
quantum level. Why do one allow a semi-classical object to
fluctuate around its classical trajectory {\it on} the surface of
SCs but not {\it perpendicular} to it? Consequently, it seems
more reasonable to relax the restrictive conditions of the RPS
quantization and only demand that in the classical limit the
trajectory of the quantized system coincides with the classical
trajectory.\par One identifies a system classically with a number
of coordinates $q_i(t)$s where $t$ is the parameter of the
evolution, the time. In principle if $q_i(t)$'s are known we know
every thing about the system. Dynamic of the system is given by
the principle of least action or equivalently by Euler Lagrange
equations of motion. In quantum mechanics one deals with
coordinate operators $\hat{q}(t)$s which act on a definite
Hilbert space. Dynamic is given by a time evolution operator.
Given a system with classical trajectory $q(t)$, we say it is
quantum mechanically described by operators $\hat{q}(t)$s if and
only if in the classical limit, $\left<\hat{q}(t)\right>=q(t)$,
where $\left<\hat{q}(t)\right>$ is the expectation value of
$\hat{q}(t)$. We call this method, generalized Gupta-Bleuler
(GGB) quantization. In the GGB method one deals with operators
$\hat{q}(t)$s which are quantum versions of coordinates $q(t)$s.
Conjugate to each $\hat{q}$ one defines an operator $\hat{p}$
which is the generator of translation in the $q$ representation
of the Hilbert space, i.e.
$\left<q|\hat{p}|\psi\right>=-i\hbar\frac{\partial}{\partial
q}\left<q|\psi\right>$ or equivalently
$\left[\hat{q},\hat{p}\right]=i\hbar$. The operator $\hat{p}$ is
not considered to be a quantum version of the momentum
$p=\frac{\partial}{\partial \dot{q}}L$. One constructs a
Hamiltonian $\hat{H}=\hat{H}(\hat{q},\hat{p})$ and gives the
dynamic by Schr\"{o}dinger equation or by Heisenberg equation. In
general there is no relation between operators $\hat{p}$ and
$\hat{H}$ and classical quantities $p=\frac{\partial}{\partial
\dot{q}}L$ and $H=p\dot{q}-L$. Quantum mechanics gives
$\left<\hat{q}(t)\right>$ and Euler Lagrange equations give
$q(t)$. One only demands that the equality
$\left<\hat{q}(t)\right>=q(t)$ be satisfied in the classical
limit. The following example exhibits the equivalence between the
GGB quantization and Schr\"{o}dinger approach in the case of
nonsingular Lagrangians. In fact this is the Ehrenfest theorem
reversed.
\par Consider a system given by the Lagrangian,
\begin{equation}\label{v1}
L(q,\dot{q})=\frac{1}{2}\dot{q}^2-V(q).
\end{equation}
The Euler Lagrange equation of motion is
\begin{equation}
\ddot{q}+\frac{\partial}{\partial q}V(q)=0.
\end{equation}
In the GGB quantization one says this system is quantum
mechanically described by the operator $\hat{q}(t)$ if in the
Heisenberg picture $\hat{q}$ satisfies the same equation of motion
\begin{equation}
\ddot{\hat{q}}+\frac{\partial}{\partial \hat{q}}V(\hat{q})=0.
\end{equation}
This can be achieved if one defines the Hamiltonian
$\hat{H}=\frac{1}{2}\hat{p}^2+V(\hat{q})$, where $\hat{p}$ is the
generator of translation in the $q$-representation of the Hilbert
space i.e. $[\hat{q},\hat{p}]=i\hbar$. In the next section we
show that this Hamiltonian could be uniquely determined by
considering two conditions. The equations of motion are given by
the Heisenberg equations,
\begin{equation}\begin{array}{cc}
\dot{\hat{q}}=\frac{-i}{\hbar}[\hat{q},\hat{H}],& \dot{\hat{p}}=
\frac{-i}{\hbar}[\hat{p},\hat{H}].
\end{array}\end{equation}
It is very important to note that $\hat{p}$ is not the operator
version of the momentum $p=\frac{\partial}{\partial \dot{q}}L$.
The operator $\hat{p}$ is only the generator of translation in the
$q$-representation of the Hilbert space,
$\left<q|p|\psi\right>=-i\hbar\frac{\partial}{\partial
q}\left<q|\psi\right>$. The above observed similarities between
the operators $\hat{p}$ and $\hat{H}$ and the classical
quantities $p=\frac{\partial}{\partial \dot{q}}L$ and
$H=p\dot{q}-L$ are only due to the particular form of the
Lagrangian (\ref{v1}) which is quadratic with respect to the
velocity. Most of the Lagrangians used to formulate ordinary
classical models are in this form.
\section{Reduced Hilbert Space in the GGB Quantization}
Although the GGB quantization is equivalent to the RPS
quantization (Schr\"{o}dinger approach) when it is applied to
unconstrained systems given by regular Lagrangian but results
could be completely different for constraint systems. As an
example consider the Lagrangian
\begin{equation}\label{a1}
L=\dot{x}\dot{y}+yz,
\end{equation}
which leads to the following ELs
\begin{eqnarray}\label{a2}\begin{array}{cc}
\ddot{x}-z=0,& y=0.
\end{array}\end{eqnarray}
Since $z$ remains undetermined, given $z=E(t)$ one can interpret
these equations to be equations of motion for a one dimensional
charged particle subjected to an external electric field, ${\bf
E}=E(t){\bf e_x}$. One can easily verify that the RPS quantization
leads to no dynamic for this system because it is given by a
singular Lagrangian and the corresponding reduced phase space is
null. In the GGB quantization one gives dynamic by the Hamitonian
\begin{equation}
\label{a3} \hat{H}=\frac{1}{2}\hat{p}_x^2-xz+\dot{E}(t)\hat{p}(z).
\end{equation}
where $\hat{p}_x$ and $\hat{p}_z$ are generators of translation
in the corresponding directions. The Hamiltonian (\ref{a3}) is
uniquely determined regarding two conditions:
\par
$\ i)$ Considering Els (\ref{a2}), the Hamiltonian should be
quadratic in $\hat{p}_x$ and linear in all other momenta. It
should contain kinetic terms corresponding to coordinates which
their accelerations appear in ELs. No kinetic term corresponding
to other coordinates that their accelerations are not appeared  in
ELs should be included in the Hamiltonian.
\par
$ii)$ This Hamiltonian leads to Heisenberg equations equivalent
to ELs in which coordinates $x_i$ are replaced by operators
$\hat{x}_i$.
\par
We do not consider the coordinate $y$ in quantization similar to
the RPS quantization. Consequently the GGB quantization leads to
a dynamic that coincides with the classical dynamic in the
classical limit. This is an important result but it is not
sufficient yet. The Hamiltonian (\ref{a3}) is defined on a two
dimensional Hilbert space expanded for example by
$\left|x\right>\left|z\right>$. Since we are studying a one
dimensional system, the Hilbert space should be one dimensional.
Considering Eq.(\ref{a2}), $x$ is the spatial coordinate and $z$
is only an auxiliary coordinate playing role of external electric
field. Consequently, physical Hilbert space is the subspace
expanded by $\left|x\right>$. How can one reduce the Hilbert
space $\left|x\right>\left|z\right>$ to realize the true subspace
$\left|x\right>$? Quantum invariants \cite{Ries,Gao} provide a
suitable answer. A Hermitian operator $\hat{I}(t)$ is called an
invariant if it satisfies
\begin{equation}\label{a4}
\frac{\partial \hat{I}}{\partial
t}-\frac{i}{\hbar}\left[\hat{I},\hat{H}\right]=0.
\end{equation}
Considering the eigen value equation of $\hat{I}(t)$
\begin{equation}
\hat{I}(t)\left|\lambda_n,t\right>=\lambda_n\left|\lambda_n,t\right>,
\end{equation}
one can show that the eigen values $\lambda_n$ are independent of
time and the particular solution $\left|\lambda_n,t\right>_S$ of
the Schr\"{o}dinger equation
$\hat{H}\left|\psi\right>_S=i\hbar\frac{\partial}{\partial
t}\left|\psi\right>_S$ is different from the eigen state
$\left|\lambda_n,t\right>$ of $\hat{I}(t)$ only by a phase factor,
$\left|\lambda_n,t\right>_S=\exp(i\phi_n(t))\left|\lambda_n,t\right>$.
Given the Hamiltonian (\ref{a3}), an invariant is
\begin{equation}
\hat{I}(t)=\hat{z}-E(t).
\end{equation}
The null eigen function of this invariant is
\begin{equation}
\psi(x,P_z)=\exp\left[-\frac{i}{\hbar}E(t)p_z\right]\phi(x).
\end{equation}
Since
\begin{equation}
\hat{H}\psi(x,p_z)=\exp\left[-\frac{i}{\hbar}E(t)p_z\right]
\left(\frac{1}{2}\hat{p}_x^2-E(t)\hat{x}+\dot{E}(t)p_z\right)\phi(x)
\end{equation}
and
\begin{equation}
i\hbar\frac{\partial}{\partial t}
\psi(x,p_z)=\exp\left[-\frac{i}{\hbar}E(t)p_z\right]
\left(\dot{E}(t)p_z+i\hbar\frac{\partial}{\partial
t}\right)\phi(x),
\end{equation}
the Schr\"{o}dinger equation becomes effectively
\begin{equation}\label{a5}
\hat{H}_e\phi(x)=i\hbar\frac{\partial}{\partial t}\phi(x),
\end{equation}
where $\hat{H}_e$ is the effective Hamiltonian
\begin{equation}
\hat{H}=\frac{1}{2}\hat{p}_x^2-E(t)\hat{x}.
\end{equation}
It is interesting to note that if one gives the classical dynamic
of the charged particle by the regular Lagrangian
$L=\frac{1}{2}\dot{x}^2+E(t)x$ instead of the singular Lagrangian
(\ref{a1}), Schr\"{o}dinger approah leads to the same
Schr\"{o}dinger equation Eq.(\ref{a5}).\par Quantum invariants
provide a suitable framework for reducing Hilbert space. Physical
states are (null) eigen states of the quantum invariant
$\hat{I}(t)$. This is a familiar statement. In the Dirac
quantization, one defines physical states as null eigen states of
the generator of gauge transformation $G$ \cite{Henbook},
\begin{equation}
\hat{G}\left|Phys\right>=0.
\end{equation}
It is shown that $G$ satisfies the following condition
\cite{Pons1,Pons2},
\begin{equation}\label{a6}
\frac{\partial}{\partial t}G+\left\{G,H\right\}=PC,
\end{equation}
where $PC$ stands for any linear combination of primary
constraints and $H$ is the classical Hamiltonian. Comparing
Eq.(\ref{a6}) with Eq.(\ref{a4}), one verifies that on the surface
of primary constraints, $G$ is an invariant.\par As one further
example, we study briefly the formulation of a charged time
dependent one dimensional oscillator coupled to an external
electric field. Consider the Lagrangian
\begin{equation}
L=\dot{q}\dot{y}-qz\dot{y}-\dot{q}zy+qz^2y+xy,
\end{equation}
that leads to the following equations of motion
\begin{eqnarray}\label{a7}
&&\ddot{q}-q(\dot{z}+z^2)-x=0,\\
&&y=0.
\end{eqnarray}
The coordinates $x$ and $z$ remain arbitrary functions of time.
Given $x=E(t)$ and $z=A(t)$ one verifies that Eq.(\ref{a7}) can
be equivalently obtained from the regular Lagrangian
\begin{equation}\label{a8}
L=\frac{1}{2}\left(\dot{q}-A(t)q\right)^2+E(t)q.
\end{equation}
For quantization we introduce operators $\hat{q}$, $\hat{x}$,
$\hat{z}$ and the corresponding generators of translation
$\hat{p}$, $\hat{p}_x$ and $\hat{p}_z$. Again we do not consider
$y$ in quantization. Considering Eq.(\ref{a7}) the Hamiltonian
should be a sum of the kinetic term $\frac{1}{2}\hat{p}^2$ and a
linear combination of other momenta. In addition, Heisenberg
equations of motion should be equivalent to Eq.(\ref{a7}) written
for operators. A Hamiltonian (maybe our unique choice) is
\begin{equation}
\hat{H}=\frac{1}{2}\hat{p}^2-\hat{q}\hat{x}+\frac{1}{2}(\hat{q}\hat{p}+\hat{p}\hat{q})\hat{z}+
\dot{A}(t)\hat{p}_x+\dot{E}(t)\hat{p}_z.
\end{equation}
Quantum invariants are
\begin{equation}\begin{array}{cc}
\hat{I}_1=\hat{x}-E(t),&\hat{I}_2(t)=\hat{z}-A(t).
\end{array}\end{equation}
Physical states are simultaneous (null) eigen states of these
invariants,
\begin{equation}
\psi(q,p_x,p_z)=\exp\left[-\frac{i}{\hbar}E(t)p_x\right]
\exp\left[-\frac{i}{\hbar}A(t)p_z\right]\phi(q).
\end{equation}
Consequently Schr\"{o}dinger equation could be effectively given
as follows,
\begin{equation}
\hat{H}_e\phi(q)=i\hbar\frac{\partial}{\partial t}\phi(q),
\end{equation}
where
\begin{equation}
\hat{H}_e=\frac{1}{2}\left(\hat{p}+\hat{q}A(t)\right)^2-\hat{q}E(t)-\frac{1}{2}\hat{q}^2A^2(t).
\end{equation}
This is what we expected from Eq.(\ref{a8}).\par Although the
above two models are very simple, but they reflect the main
aspects of the method. One can show that for familiar gauge field
theories like QED, the GGB quantization is completely equivalent
to the RPS quantization. This result is a direct consequence of
simple constraint-structures in these models. In general,
classification of Lagrangians for which the GGB quantization and
the RPS quantization are equivalent is not trivial and should be
studied.
\section{Conclusions} A system
is identified by its dynamic given classically by Euler Lagrange
equations of motion. Given a Lagrangian, there may exist other
Lagrangians, particularly singular Lagrangians, that lead to
equivalent dynamic. Consequently, quantization should be
independent of the particular form of the Lagrangian.
Quantization approach should be formulated such that in the
classical limit quantum mechanic lead to classical dynamic given
by Euler Lagrange equations of motion. The generalized
Gupta-Bleuler quantization is based on these considerations. We
showed that this method is equivalent to Schr\"{o}dinger method
for unconstrained Lagrangians. In the case of constraint systems
we defined the reduced Hilbert space which is the true physical
subspace of the Hilbert space by using the concept of quantum
invariants. We studied two simple systems given by singular
Lagrangians and showed that the generalized Gupta-Bleuler
quantization leads to true Schr\"{o}dinger equations.
\par {\bf Acknowledgement}\par The author would like to
thank K. Samani for useful discussions and comments.


\newpage
\begin{thebibliography}{99}

\bibitem{Dirac} P.A. M. Dirac, Can. J. Math. {\bf 2}, (1950) 129 ;
Proc. R. Soc. London Ser. A {\bf 246}, (1958) 326; {\it "Lectures
on Quantum Mechanics"} New York: Yeshiva University Press, 1964,
\bibitem{Berg} J. L. Anderson and P. G. Bergman, Phys Rev. {\bf
83}, (1951) 1018.
\bibitem{Henbook} M. Henneaux and C. Teitelboim
{\it "Quantization of Gauge System"}, Princeton University Press, Princeton, New
Jersey, 1992.
\bibitem{Itz} C. Itzykson and J. B. Zuber, {\it "Quantum Field
Theory"}, McGraw-Hill Inc., 1980.
\bibitem{Gitman} D. M. Gitman and I. V. Tyutin, {\it Quantization of Fields with
Constraints}, Springer-Verlag Berlin Heidelberg, 1990.
\bibitem{Gov} J. Govaerts, {\it Hamiltonian Quantization and
Constraines Dynamics}, Leuven Notes in Mathematical and
Theoretical Physics, Vol. 4, 1991.
\bibitem{Klauder} J. R. Klauder, Lect.Notes Phys. {\bf 572}
(2001) 143.
\bibitem{Pons} C. Batlle, J. Gomis, J. M. Pons, N. Roman Roy, J.
Math. Phys. {\bf 27} (12), (1986) 2953.
\bibitem{Ries} H. R. Lewis and W. B. Riesenfeld, J. Math. Phys. {\bf 10}, (1969) 1458.
\bibitem{Gao} X. C. Gao, J. B. Xu and T. Z. Qian, Phys. Rev. A
{\bf 44}, (1991) 7016.
\bibitem{Pons1} C. Batlle, J. Gomis, X. Gracia and J. M. Pons, J.
Math. Phys. {\bf 30} (6), (1986) 1345.
\bibitem{Pons2} J. M. Pons and J. A. Garcia, Int. J. Mod. Phys. A {\bf 15}
(2000) 4681.
\end{thebibliography}
\end{document}